\newcommand{\CLASSINPUTtoptextmargin}{1.5cm}
\newcommand{\CLASSINPUTbottomtextmargin}{1.5cm}
\documentclass[conference]{IEEEtran}
\usepackage{cite}
\usepackage{amsmath,amssymb,amsfonts}
\usepackage{algorithmic}
\usepackage{graphicx}
\usepackage{textcomp}
\usepackage{xcolor}
\usepackage{hyperref}
\usepackage{booktabs}
 \IEEEoverridecommandlockouts
\begin{document}
\title{Robust Model Predictive Techno-Economic Control of Active Distribution Networks\vspace{-0.1in}}
\author{\IEEEauthorblockN{Salish Maharjan, Prashant Tiwari, Rui Cheng, Zhaoyu Wang}
\IEEEauthorblockA{\textit{Iowa State University, Iowa, U.S.A.}\\
salish@iastate.edu, ptiwari@iastate.edu, ruicheng@iastate.edu, and wzy@iastate.edu}\vspace{-0.35in}
\thanks{\scriptsize{This work was supported in part by the National Science Foundation under ECCS 1929975, ECCS 2042314, and in part by the U.S. Department of Energy Wind Energy Technologies Office under Grant DE-EE0008956.}}}

\maketitle

\begin{abstract}
Stochastic controllers are perceived as a promising solution for techno-economic operation of distribution networks having higher generation uncertainties at large penetration of renewables. These controllers are supported by forecasters capable of predicting generation uncertainty by means of lower/upper bounds rather than by probability density function (PDF). Hence, the stochastic controller assumes a suitable PDF for scenario creation and optimization, requiring validation of the assumption. To effectively bridge the forecaster’s capability and resolve the assumption issues, the paper proposes a robust model prediction-based techno-economic controller, which essentially utilizes only the lower/upper bounds of the forecast, eliminating the necessity of PDF. Both discrete and continuous control resources such as tap-changers and DERs are utilized for regulating the lower/upper bounds of the network states and robustly minimizing the cost of energy import. The proposed controller is implemented for UKGDS network and validated by comparing performance at various confidence levels of lower/upper bound forecast.
\end{abstract}


\section{Introduction}
The growing integration of renewable Distributed Energy Resources (DERs) has devolved the distribution system with many technical challenges (e.g., regulating voltage, line current, and reverse power flow) and evolved opportunities for more economical operation (e.g., optimizing the energy import and delivering ancillary services). Resolving technical challenges and achieving cost-efficient operation (referred to as techno-economic operation) is the desirable objective of network controllers at high penetration of renewable DERs; however, their large uncertainties oppress the objective by inducing uncertainty to network states (e.g., voltage and line current). In literature, the techno-economic operation of the distribution networks is studied considering the uncertainty of renewable DERs as a day-ahead optimization problem \cite{Gao2018IntegratedSystem, Xiong2021ANetworks}, multi-timescale control problem \cite{Xu2017Multi-TimescaleSystems, Zafar2018Multi-timescaleGrids}, and single-timescale control problem \cite{Jiang2019StochasticRenewables, Maharjan2021RobustNetworks}. 
\par 
A day-ahead scheduling-based techno-economic operation is usually popular in active distribution networks that do not have advanced communication infrastructure. Integrated two-stage stochastic scheduling is studied in \cite{Gao2018IntegratedSystem} where the slow-acting (switches)/time-coupled(storage system) and all fast-acting decision variables are optimized in the first and second stages respectively, considering the uncertainty of renewable DERs. A similar problem is studied in \cite{Xiong2021ANetworks} using a single-stage chance-constrained optimization by employing only the key scenarios generated by grouping probability density functions (PDF) of renewable DERs. The day-head stochastic scheduling method is prone to deviate from optimal operation at a higher degree of uncertainty and during unplanned contingencies. Hence closed-loop controllers  \cite{Xu2017Multi-TimescaleSystems, Zafar2018Multi-timescaleGrids,Jiang2019StochasticRenewables, Maharjan2021RobustNetworks} are generally preferred for techno-economic optimization, albeit they demand complex communication infrastructure. 
\par
Stochastic closed-loop techno-economic control is designed as a recourse action \cite{Xu2017Multi-TimescaleSystems, Zafar2018Multi-timescaleGrids} by the continuous acquisition of measurements and optimization of fast-acting decision variables over multiple scenarios, whereas slow-acting controllers are optimized in slower timescale. Although multi-timescale techno-economic controls are efficient, their implementation would require both short/long-term forecasters, and hence may not be cost-effective. In contrast, single-timescale control requires only a short-term forecaster, which is designed using the concept of model predictive control (MPC) for techno-economic operation in \cite{Jiang2019StochasticRenewables, Maharjan2021RobustNetworks}. As MPC could be designed for various control horizons, they could be optimally designed for controlling both slow/fast-acting controllers in one timescale. A stochastic MPC-based techno-economic controller is designed in \cite{Jiang2019StochasticRenewables} considering a normal distribution of generation/load uncertainty. However, their PDF of uncertainty for the short-term mainly for renewables is not certainly known and is debatable as the forecasters essentially forecast the lower/upper bounds (prediction intervals) not the PDF \cite{Ni2017AnForecasting}. This issue is highlighted in \cite{Maharjan2021RobustNetworks}, and  hence it proposes a robust MPC, which does not require scenario generation using the PDF. The robust MPC in  \cite{Maharjan2021RobustNetworks} utilizes only the prediction interval for robust voltage regulations. 

\par
Hence, this paper extends the concept of robust MPC for techno-economic control in the distribution network considering the uncertainties in renewable DERs (essentially PVs) and load demands. Both the discrete (OLTC) and continuous (P/Q injections from renewable DERs and Battery Energy Storage (BES)) resources are optimized in a single timescale not only to robustly regulate node voltage, line currents, and reverse power flow but also robustly minimize the cost of active power import from the external grid with variable tariff. This is accomplished by using the sensitivity-based linear model of the distribution network, which can integrate both discrete and continuous variables to form a mixed-integer convex problem. The proposed RMPC is implemented to achieve closed-loop control of the UKGDS network. The closed-loop control is designed by co-simulation of a python-based RMPC controller and RMS model of UKGDS in DigSILENT PowerFactory. Additionally, the performance of RMPC is evaluated for the prediction interval forecast at various confidence levels. 

\section{Modeling of Active Distribution Network}
\subsection{Linear Model of Distribution Network}
The linear model of the distribution network is developed to characterize the network response to changes in control inputs ($\Delta U$) and disturbances ($\Delta D$) as:
\begin{small}
\begin{align}
&\textbf{V}(t) = \textbf{V}(t-1)+\hspace{-5pt}\sum_{j\in\mathcal{I}^U}\hspace{-2pt}\left[\frac{\partial \textbf{V}}{\partial U_j} \right] \Delta U_j(t) + \hspace{-5pt}\sum_{j\in\mathcal{I}^{D}} \hspace{-2pt}\left[\frac{\partial \textbf{V}}{\partial D_j} \right] \Delta D_j(t)\label{eqn_dist_lin_Vmod_ch5}\\
&\textbf{I}(t) = \textbf{I}(t-1)+\hspace{-5pt}\sum_{j\in\mathcal{I}^{U}}\hspace{-2pt}\left[\frac{\partial \textbf{I}}{\partial U_j} \right] \Delta U_j(t) + \hspace{-5pt}\sum_{j\in\mathcal{I}^{D}}\hspace{-2pt} \left[\frac{\partial \textbf{I}}{\partial D_j} \right] \Delta D_j(t) \label{eqn_dist_lin_Imod_ch5}\\
&\textbf{L}(t) = \textbf{L}(t-1)+\hspace{-5pt}\sum_{j\in\mathcal{I}^{U}}\hspace{-2pt}\left[\frac{\partial \textbf{L}}{\partial U_j} \right] \Delta U_j(t) + \hspace{-5pt}\sum_{j\in\mathcal{I}^{D}} \hspace{-2pt}\left[\frac{\partial \textbf{L}}{\partial D_j} \right] \Delta D_j(t)\label{eqn_dist_lin_Lmod_ch5}
\end{align}
\end{small}
The node voltages ($\textbf{V}$), line currents ($\textbf{I}$), and network losses ($\textbf{L}$) are projected along future time horizon ($t\in\mathcal{T}$), with reference to current measurements (i.e., at time $t-1$) in (\ref{eqn_dist_lin_Vmod_ch5}), (\ref{eqn_dist_lin_Imod_ch5}), and (\ref{eqn_dist_lin_Lmod_ch5}) respectively. The projections are made using the network's sensitivity matrices for \textbf{V}, \textbf{I}, and \textbf{L} with respect to control ($U$) and disturbances variables ($D$), estimated using enhanced Z-bus method \cite{Maharjan2020EnhancedNetworks}. The control resources comprise reactive power injections from curtaiable/non-curtailable PV sources ($[Q_j|_{j\in \mathcal{I}^{PV}}^T,Q_j|_{j\in \mathcal{I}^{CPV}}]$), power curtailment of curtailable PV sources ($P_j^{max}|_{j\in \mathcal{I}^{CPV}}$), active/reactive power injection from BES ($[P_j|_{j\in \mathcal{I}^{ES}}, Q_j|_{j\in \mathcal{I}^{ES}}]^T$), and discrete control variables of tap-changers ($ tap_{ij}|i,j\in\mathcal{I}^{OLTC}$). Whereas PV power generation ($\left[P_j|_{j\in \mathcal{I}^{PV}}, P_j|_{j\in \mathcal{I}^{CPV}}\right]$) and loads ($\left[P_j|_{j\in \mathcal{I}^{L}}, Q_j|_{j\in \mathcal{I}^{L}} \right]^T$) are treated as stochastic disturbances whose upper and lower bounds are provided by forecasted prediction intervals. The loss matrix in (\ref{eqn_dist_lin_Lmod_ch5}) comprises of active and reactive power loss, i.e., $\mathbf{L}=[P_{loss}, Q_{loss}]^T$.

\subsection{Model for PV sources and loads}
The PVs and loads are modeled by a time series obtained from a short-term forecaster, capable of predicting the lower/upper bounds ($\hat{p}^{lo}/\hat{p}^{up}$) and the average values ($\hat{p}^{ave}$). The PV short-term forecaster is assisted with sky-camera and satellite weather data for generating accurate prediction intervals as explained in \cite{Ni2017AnForecasting}, and are modeled as:
\begin{small}
\begin{align}
    P_j^{PV}(t) =& f(\hat{p}_j^{lo},\hat{p}_j^{up},\hat{p}_j^{ave},t)\quad j\in(\mathcal{I}^{PV}\cup\mathcal{I}^{CPV}), \;t\in\mathcal{T}
\end{align}
\end{small}
We assume, the non-curtailable PVs are privately owned and comply with IEEE 1547-2018 standard \cite{IEEEStandardAssociation2018IEEEInterfaces}. Hence, their reactive power capability for inverter rating of ($S^{PV}$) is defined as:
\begin{small}
\begin{align}
    -0.44S_j^{PV} \leq Q_j^{PV}(t)\leq 0.44S_j^{PV} \quad j\in \mathcal{I}^{PV},\;t\in\mathcal{T} \label{eqn_ch5_Qpv}
\end{align}
\end{small}
We assume the curtailable PVs are owned by DSO, who could utilize the full Q-margin of curtailable PV inverters whenever the active power generation is lower than the rated power ($S^{PV}$). If $P_j^{max}\in[0, \hat{p}^{up}_j]$ is the maximum power set point of curtailable PV ($j\in\mathcal{I}^{CPV}$), then the corresponding reactive power margin is defined as:
\begin{small}
\begin{align}
    (Q_j^{PV}(t))^2 \leq (S_j^{PV})^2-(P_j^{max}(t))^2,\; j\in \mathcal{I}^{CPV}, t\in\mathcal{T} \label{eqn_ch5_Qcurtpv}
\end{align}
\end{small}
Similar to non-curtailable PV resources, the loads are modeled by predicted time series defining their demand over a horizon ($\mathcal{T}$) as:
\begin{small}
\begin{align}
    P_j^{L}(t) = & f(\hat{p}_j^{lo},\hat{p}_j^{up},\hat{p}_j^{ave}, t)\quad j\in\mathcal{I}^{L},\;t\in\mathcal{T}
\end{align}
\end{small}

\subsection{Model for battery energy storage (BES)}
During the charging and discharging process, the State of Charge ($SoC$) dynamics is defined as:

\begin{small}
\begin{align}
    SoC(t) = SoC(t-1)+\frac{\eta \Delta T P^{ES}(t)}{B_{cap}}. \label{eqn_ch5_soc}
\end{align}
\end{small}
Here {\small $B_{cap}$}, {\small $P^{ES}$}, and {\small $\Delta T$} are the battery capacity (in MWhr), power exchanged from BES, and time-step, respectively. The efficiency ({\small $\eta$}) is defined as:

\begin{small}
\begin{align}
    \eta = \begin{cases}
    1/\eta^d, \quad \text{if $P^{ES}>0$ (discharging mode)}\\
    \eta^c, \quad \text{otherwise (charging mode)}
    \end{cases} \label{eqn_if_else}
\end{align}
\end{small}
The charging and discharging efficiencies ($\eta^c$ and $\eta^d$) account for the losses during respective events. {\small $P^{ES}$} is defined as positive for discharging power and negative for charging power. The standard approach to convert the  if/else condition (\ref{eqn_if_else}) into a set of mixed integer linear constraints by introducing an auxiliary variable $z(t)= \delta(t) P^{ES}(t)$ and binary variable $\delta(t)$  is detailed in \cite{Maharjan2020ANALYSISRESOURCES}, using which (\ref{eqn_ch5_soc}) and (\ref{eqn_if_else}) is reformulated with a set of linear equations defined by  (\ref{eqn_bat_SoC}) and (\ref{eqn_bat_ineq_cons}) as: 

\begin{small}
\begin{align}
    &SoC(t) = SoC(t-1)+(\eta^c-1/\eta^d)z(t)-\eta^cP^{ES}(t),\label{eqn_bat_SoC}\\
    & \mathbf{E_1}\delta(t)+\mathbf{E_2}z(t)\leq \mathbf{E_3}P^{ES}(t)+\mathbf{E_4}, \quad   \text{where} \label{eqn_bat_ineq_cons}\\
    &\mathbf{E_1} = [P^{ES,max}, -(P^{ES,max}+0.1),P^{ES,max}, P^{ES,max}]^T\\
    &\mathbf{E_2} = [0,0,1, -1, -1, -1]^T, \;\mathbf{E_3} = [1, -1, 0, 0, 1, -1]^T\\
    &\mathbf{E_4} = [P^{ES,max}, -0.1, 0, 0, P^{ES,max}, P^{ES,max}]^T
\end{align}
\end{small}

Here, {\small $P^{ES,max}$} is the maximum charge/discharge rate of BES. The SoC constraint (\ref{eqn_soc_cons}) limits the over-charging and over-discharging of the battery whereas (\ref{eqn_bat_power_const}) limits the power exchange under the safe boundary. (\ref{eqn_pow_rate_cons}) limits the maximum rate of charging/discharging ({\small $\Delta P^{ch,max} / \Delta P^{disch,max}$}). Finally, the last constraint (\ref{eqn_bat_inv_Q_margin}) is defined from IEEE 1547-2018 \cite{IEEEStandardAssociation2018IEEEInterfaces}.
\begin{small}
\begin{align}
    SoC^{lo} \leq SoC(t) \leq SoC^{up} \label{eqn_soc_cons}\\
    -P^{ES,max} \leq P^{ES}(t) \leq P^{ES,max}\label{eqn_bat_power_const}\\
    \Delta P^{ES}(t) = P^{ES}(t) - P^{ES}(t-1)\\
    -\Delta P^{ch,max}\leq \Delta P^{ES}(t)\leq \Delta P^{disch,max}\label{eqn_pow_rate_cons}\\
    -0.44S_{inv}^{ES}\leq Q^{ES}(t) \leq 0.44S_{inv}^{ES}(t)\label{eqn_bat_inv_Q_margin}
\end{align}
\end{small}

\subsubsection{Battery degradation model}
\label{sec_bat_deg_cost}
Factors that accelerate the battery aging are  temperature, high power rate, and  charging/discharging cycle of the battery. It is well known that the aging process due to all these factors is highly nonlinear, and they interact in a multiplicative way with each other \cite{Smith2012ComparisonCycles}. However, for operational optimization, the degradation cost can be modeled fairly in quadratic form by only considering the significant loss factor, which is due to the charging/discharging cycle, as in \cite{Koller2013DefiningSystem} as:

\begin{small}
\begin{align}
    J_{Bat} = J_{c/d}(SoC(t)-SoC(t-1))^2
\end{align}
\end{small}
where $J_{c/d}$ degradation cost for unit change in $SoC$.

\subsection{On-load tap changer (OLTC) }
The OLTC is commanded for unity change and is a discrete control resource. If {\small$\Delta tap(t)$} is the change in {\small$tap$} value between time {\small$t$} and {\small$t-1$}, then its operation limits are defined as:
\begin{small}
\begin{align}
    -1 \leq & \Delta tap_{ij}(t)\leq 1 \label{eqn_ch5_delTap}\\
    Tap^{min}\leq & tap_{ij}(t) \leq Tap^{max}, \quad i,j\in\mathcal{I}^{OLTC}, t\in\mathcal{T} \label{eqn_ch5_tap}
\end{align}
\end{small}
Here, the {\small$Tap^{min}$} and {\small$Tap^{max}$} are the minimum and maximum tap positions available in the OLTC, and {\small$\Delta tap_{ij}$} is an integer variable.

\section{Robust model predictive distribution network controller}
The linear model of the distribution network (\ref{eqn_dist_lin_Vmod_ch5})-(\ref{eqn_dist_lin_Lmod_ch5}) provides the evolution of node voltage, line current, and loss along the control/prediction horizon ({\small$\mathcal{T}$}) to the temporal transition of PV generations and load demands. The network controller dispatches the control set-points for achieving optimal operation of the network. However, the controller does not have exact information on PV and load transition along {\small$\mathcal{T}$}, rather it has to make an optimal decision based on their prediction intervals. Hence, the paper proposes a robust model predictive control approach for achieving robust operation utilizing the prediction intervals of disturbance variables such as PVs and loads.
\subsection{Estimation of lower/upper bounds of network states subjected to bounded disturbances}
The lower/upper bound of the network's state variable ({\small $\textbf{X}=\{\textbf{V}, \textbf{I}, \textbf{L}$\}}) subjected to the prediction interval of PVs and loads along the {\small$\mathcal{T}$} could be estimated by:
\begin{small}
\begin{align}
    \textbf{X}^{lo/up}(t) =& \textbf{X}^{lo/up}(t-1) + \mathbf{\Delta \textbf{X}}^{lo/up}(t), \label{eqn_open_loop_est} \quad \text{where}\\
    \mathbf{\Delta \textbf{X}}^{up}(t) =&\sum_{j\in\mathcal{I}^{D}} \left(\frac{\partial \textbf{X}}{\partial P_j}\right) \Bigg[ \Delta P^{max}_j(t)sgn\left(\frac{\partial \textbf{X}}{\partial P_j}\right)+\nonumber\\& \Delta P^{min}_j(t)\left\{\textbf{1}-sgn\left(\frac{\partial \textbf{X}}{\partial P_j}\right)\right\}\Bigg]\label{eqn_deltaV_pv_up_ch5}\\
   \mathbf{\Delta \textbf{X}}^{lo}(t) =&  \sum_{j\in\mathcal{I}^{D}} \left(\frac{\partial \textbf{X}}{\partial P_j}\right) \Bigg[ \Delta P^{max}_j(t)\left\{\textbf{1}-sgn\left(\frac{\partial \textbf{X}}{\partial P_j}\right)\right\}+\nonumber\\& \Delta P^{min}_j(t)sgn\left(\frac{\partial \textbf{X}}{\partial P_j}\right)\Bigg] \label{eqn_deltaV_pv_lo_ch5}
\end{align}
\end{small}
As advocated in \cite{Maharjan2021RobustNetworks, Maharjan2020ANALYSISRESOURCES}, the {\small $\Delta P^{max}_j\text{ and } \Delta P^{min}_j$} are determined using the prediction intervals [$\hat{p}_j^{lo}(t),\hat{p}_j^{up}(t)$] of disturbance variable pertaining to PVs and loads as:
\begin{small}
\begin{align}
    \Delta P^{min/max}_j(t) = \hat{p}_j^{lo/up}(t) - \hat{p}_j^{lo/up}(t-1)\quad j\in\mathcal{I}^D
\end{align}
\end{small}

\subsection{Estimation of lower/upper bounds of network states subjected to bounded disturbances and control variables}
The expression (\ref{eqn_open_loop_est}) provides the open loop response of the network to a predicted disturbance. To find the closed loop response, (\ref{eqn_open_loop_est}) is augmented with perturbation of network states due to change in control variables ({\small$\mathbf{\Delta \textbf{X}}^{ctr}$}). Unlike disturbance, the control variable is deterministic in nature. Hence, the closed-loop response of the network state is given by:
\begin{small}
    \begin{align}
        \textbf{X}^{lo/up}(t) =& \textbf{X}^{lo/up}(t-1) + \mathbf{\Delta \textbf{X}}^{lo/up}(t) + \mathbf{\Delta \textbf{X}}^{ctr}(t), \text{where} \label{eqn_closed_loop_est}\\
        \mathbf{\Delta \textbf{X}}^{ctr}(t)=& \sum_{j\in\mathcal{I}^U}\left[\frac{\partial \textbf{X}}{\partial U_j} \right] \Delta U_j(t) \\
        \Delta U_j(t) =& U_j(t) - U_j(t-1)
    \end{align}
\end{small}

\subsection{Estimation of lower/upper bound of active power demand from the external grid}
The active power from the external grid is determined using the active power balance equation as: 
\begin{small}
\begin{align}
    \sum_{j\in\mathcal{I}^{EG}}P^{EG}_j(t)=\sum_{j\in\mathcal{I}^L}P_j(t)+P_{loss}(t)-\hspace{-0.32in}\sum_{j\in(\mathcal{I}^{PV}\cup\mathcal{I}^{CPV})}\hspace{-0.32in}P_j(t)-\sum_{j\in\mathcal{I}^{ES}}\hspace{-5pt}P_j(t)
\end{align}
\end{small}
As the PV sources and loads are defined by prediction intervals, the power drawn from the external grid will also have certain uncertainty bounds. We are interested in the upper/lower limit of it and are estimated as:
\begin{small}
\begin{align}
\sum_{j\in\mathcal{I}^{EG}}P^{EG,up}_j(t)=&\sum_{j\in\mathcal{I}^L}\hat{P}^{up}_j(t)+P_{loss}^{up}(t)-\sum_{j\in\mathcal{I}^{PV}}\hat{P}^{lo}_j(t)-\nonumber\\&\hspace{-0.55in}\sum_{j\in\mathcal{I}^{CPV}}\big(\hat{P}^{lo}_j(1-\delta_j(t))+z_j(t)\big)-\sum_{j\in\mathcal{I}^{ES}}P_j(t) \label{eqn_P_EG_up}\\
    \sum_{j\in\mathcal{I}^{EG}}P^{EG,lo}_j(t)=&\sum_{j\in\mathcal{I}^L}\hat{P}^{lo}_j(t)+P_{loss}^{lo}(t)-\sum_{j\in\mathcal{I}^{PV}}\hat{P}^{up}_j(t)-\nonumber\\&\hspace{-0.55in}\sum_{j\in\mathcal{I}^{CPV}}P_j^{max}(t)-\sum_{j\in\mathcal{I}^{ES}}P_j(t) \label{eqn_P_EG_lo}
\end{align}
\end{small}
The upper/lower bounds of {\small$P_{loss}$} are given by (\ref{eqn_deltaV_pv_up_ch5}) and (\ref{eqn_deltaV_pv_lo_ch5}). As defined in \cite{Maharjan2021RobustNetworks}, the upper/lower bounds of power from curtailable PVs depend on the level of curtailment ({\small$P^{max}\in[0, \hat{P}^{up}]$}). The upper bound of curtailable PV is {\small$P_j^{max}$}, whereas the lower bound ({\small$P_j^{lo}$}) will be defined as:
\begin{small}
\begin{align}
    P^{lo}_j(t) = \begin{cases}
    \hat{P}^{lo}_j(t),\quad \text{if}\;P_j^{max}(t)\ge \hat{P}^{lo}(t)\\
    P^{max}(t), \quad \text{otherwise}.\quad \qquad j \in \mathcal{I}^{CPV}
    \end{cases} \label{eq_conditon_Plow}
\end{align}
\end{small}
Again, the  conditional expression (\ref{eq_conditon_Plow}) is reformulated with auxiliary and binary variables in (\ref{eqn_P_EG_up}).

\subsection{Robust objective function}
The objective function is defined to minimize multiple cost functions, which comprise (a) the sum of the maximum penalty for voltage and line current violations, (b) the maximum active power purchased from the external grid, (c) the cost of control resources ($\mathbf{\Delta u}$), (d) penalty of PV power curtailment, (d) cost of battery degradation and (e) cost of power exchanged from BES. Furthermore, the controller minimizes these objectives over a finite time horizon by anticipating the prediction intervals of PV sources and loads.  
\begin{small}
\begin{align}
    \mathbf{min}\sum_{t\in\mathcal{T}} \Big(&\sum_{i\in\mathcal{I}^b}\mathbf{max}\; VV\{V_i(t)\}+\sum_{i\in\mathcal{I}^{li}}\mathbf{max}\;IV\{I_i(t)\}+\nonumber\\ &\hspace{-0.55in}\sum_{j\in\mathcal{I}^{EG}}\hspace{-10pt}\mathbf{max}\;PPC\{P^{EG}_j(t)\}+CC\{\mathbf{u}(t)\}+\hspace{-10pt}\sum_{j\in\mathcal{I}^{CPV}}\hspace{-10pt}CurPC\{P^{max}_j(t)\}+\nonumber\\ &\hspace{-0.55in}\sum_{j\in\mathcal{I}^{ES}}   DC\{SoC_j(t)\}+ \sum_{j\in\mathcal{I}^{ES}}\;PESC\{P^{ES}_j(t)\}\Big) \label{eqn_ch5_rob_obj_func}
\end{align}
\end{small}
The control resources comprise of tap changes from OLTC, reactive power injection from curtailable/non-curtailable PV sources, and active/reactive power injections from BES, i.e., {\small$\mathbf{\Delta u} = [\Delta\mathbf{tap}, \Delta\mathbf{Q}^{PV}, \Delta\mathbf{Q}^{CPV}, \Delta\mathbf{Q}^{ES}]$}. The active power exchange from BES and PV active power curtailment are other control resources whose cost functions are separately addressed by {\small$PESC \text{ and } CurPC$} in (\ref{eqn_ch5_rob_obj_func}).
\begin{figure}
    \centering
    \includegraphics[width=0.8\linewidth]{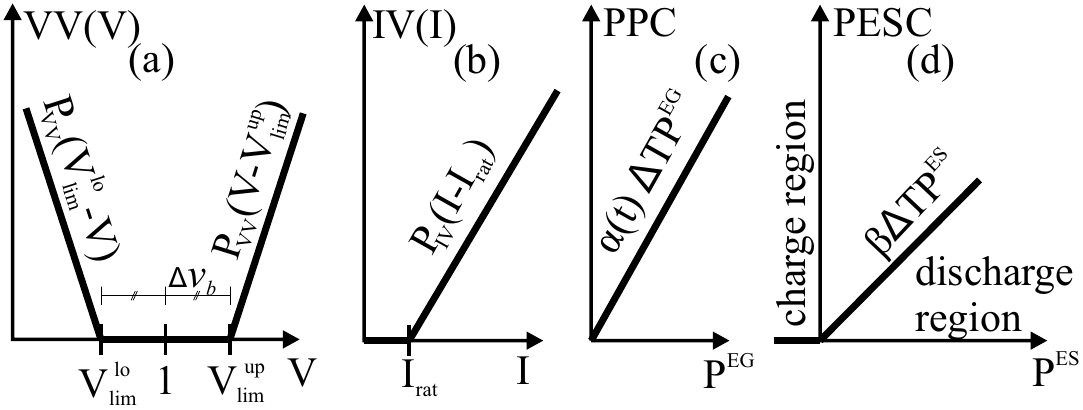}
    \vspace{-10pt}
    \caption{Penalty function imposed against (a) voltage violation and (b) current limit violation. The cost function for  (c) power import (d) power exchanged from  BES.}
    \label{fig_penalty_cst_func}
    \vspace{-15pt}
\end{figure}

\subsubsection{Voltage and line current violation penalty (\texorpdfstring{$VV\{\cdot\}$}{} and \texorpdfstring{$IV\{\cdot\}$}{})}
The node voltages and line currents are required to be kept under a targetted limit $[V_{lim}^{lo},V_{lim}^{up}]$ and $[0,I_{rat}]$. These constraints have been defined by introducing a violation cost which is the set of affine functions (shown in Fig. \ref{fig_penalty_cst_func}(a),(b)). 

\subsubsection{Active power purchase cost \texorpdfstring{$(PPC\{\cdot\})$}{}}
The DSO procures the demanded power from the wholesale market. The unit cost of electricity in the wholesale market is updated every half an hour \cite{EnergyMarketCompany2020PriceInformation}. The DSO might inject the active power to the HV external grid if it is contracted or under coordinated dispatch \cite{Zheng2018OptimalUncertainty}. In this paper, we assume the DSO is not allowed to inject active power into the HV grid. It means the excess power has to be either curtailed or absorbed in case of surplus power generation from distributed PVs. If $P_j^{EG}$ is the active power demanded from the external grid, the purchase cost is defined by a linear function shown in Fig. \ref{fig_penalty_cst_func}(c). 
Here $\alpha(t)$ are the prices of active power at the wholesale market which is defined in terms of energy. 

\subsubsection{Control Cost \texorpdfstring{($CC\{\cdot\})$}{}}
The control cost is defined as {\small $CC\{\mathbf{u}\} = \mathbf{\Delta u}^T R \mathbf{\Delta u}$. Here, $\mathbf{\Delta u} = [\Delta tap, \Delta \mathbf{Q}^{PV}, \Delta \mathbf{Q}^{CPV},\Delta \mathbf{Q}^{ES}]^T$. $R$} is the diagonal matrix comprising of penalty for changes in control resources. 

\subsubsection{PV curtailment penalty cost \texorpdfstring{$(CurPC\{\cdot\})$}{}}
The lower and upper bound of the PV generation is provided by the forecasting methods. The curtailable PVs could be operated below the upper bound as required; however, this provision is suppressed by the following penalty function.

\begin{small}
\begin{align}
    CurPC\{P^{cur}_j\} = P_{PC}(\hat{P}^{up}_j-P^{max}_j)\label{eqn_CC}\quad \quad j\in\mathcal{I}^{CPV}
\end{align}
\end{small}
Here $P_{PC}$ is the penalty factor for active power curtailment from curtailable PVs.
\subsubsection{Battery Degradation Cost \texorpdfstring{$(DC\{\cdot\})$}{}}
The details of battery degradation are modeled in section \ref{sec_bat_deg_cost}. 
\subsubsection{Cost of Power Exchange from BESS \texorpdfstring{$(PESC\{\cdot\})$}{}}
The battery is discharged only to prevent voltage violations and during the period of peak wholesale price of electricity. To favor the proposed conditions, the cost of power exchange from BES has been defined separately for charging and discharging status with the help of the affine function shown in Fig. \ref{fig_penalty_cst_func}(d) where $\beta$ is the unit cost of energy discharged.
\subsection{Max removal transformation in robust objective function}
The objective involved \textbf{min-max} terms in (\ref{eqn_ch5_rob_obj_func}), which is converted to a minimization problem by introducing auxiliary variables $\upsilon$, $\epsilon$, and $\gamma$ for voltage violation penalty, line current violation penalty, and power purchased cost from external grid respectively. The subsequent objective after removing the \textbf{max} operator is shown in (\ref{eqn_max_removed_obj}), which is equivalent to the original objective (\ref{eqn_ch5_rob_obj_func}) when constrained with additional constraints listed from (\ref{eqn_vv1})-(\ref{eqn_ppg2}).

\begin{small}
\begin{align}
    &\mathbf{min}\sum_{t\in\mathcal{T}} \Big(\sum_{i\in\mathcal{I}^b}\upsilon_i(t)+\sum_{i\in\mathcal{I}^{li}}\epsilon _i(t)+\sum_{j\in\mathcal{I}^{EG}}\gamma_i(t)+CC\{\mathbf{u}(t)\}+\nonumber\\ &\hspace{-8pt}\sum_{j\in\mathcal{I}^{CPV}}\hspace{-10pt}CurPC\{P^{max}_j(t)\}+\hspace{-7pt}\sum_{j\in\mathcal{I}^{ES}}\hspace{-8pt}   DC\{SoC_j(t)\}+\hspace{-8pt} \sum_{j\in\mathcal{I}^{ES}}\hspace{-10pt}PESC\{P^{ES}_j(t)\}\Big) \nonumber\\
    &\text{subject to:}\label{eqn_max_removed_obj}\\
    &\upsilon_i \ge P_{VV}(V_{tar}^{lo}-V^{lo}_i),\:
    \upsilon_i \ge P_{VV}(V^{up}_i-V_{tar}^{up}),
    \upsilon_i \ge 0, i\in\mathcal{I}^b\label{eqn_vv1}\\
    &\epsilon_i \ge P_{IV}(I^{up}_i-I_{rat}),\; \text{and} \quad
    \epsilon_i \ge 0.\quad \qquad i\in\mathcal{I}^{li} \label{eqn_iv2}\\
    &\gamma_j \ge \alpha(t) \Delta T P^{EG,up}_j, \qquad \qquad j\in\mathcal{I}^{EG} \label{eqn_ppg2}
\end{align}
\end{small}
The resulting optimization problem results in the form of Quadratic Constraint Mixed-Integer Quadratic Programming (QCMIQCP).

\section{Simulation Results}
The UKGDS network is considered for study (refer to Fig. 2 in \cite{Maharjan2019IntegrationDevices}). It has 22 distributed PVs whose aggregated installed capacity is 80\% of the substation capacity (52.8 MVA). The additional DER introduced in this study is the BES which is connected to node \textbf{1100}. The BES has an inverter capacity of 15 MVA with a battery capacity of 100 MWh. The proposed RMPC is implemented to achieve closed-loop control of the UKGDS network. The closed-loop control is designed by co-simulation of a python-based RMPC controller and RMS model of UKGDS in DigSILENT PowerFactory. The PV and load prediction interval are considered  with 100\% CL unless specified in the case studies presented below, and their daily base profile is adopted from \cite{Maharjan2021RobustNetworks}. The prediction horizon of RMPC is taken to be an hour which comprises four time-steps of 15 minutes. All required parameters for the presented simulation results are listed in Table \ref{tab_para_rcmpc_ctrl}.
\begin{table}[t]
  \centering
  \caption{Parameters of RCMPC based controller}
    \begin{tabular}{cccccc}
    \toprule
    \multicolumn{1}{l}{\textbf{Item}} & \multicolumn{1}{l}{\textbf{Value}} & \multicolumn{1}{l}{\textbf{Item}} & \multicolumn{1}{l}{\textbf{Value}} &  \multicolumn{1}{l}{\textbf{Item}} & \multicolumn{1}{l}{\textbf{Value}}\\
    \midrule
    $P_{VV}$ & 1000000\$/$pu^2$ & $N_p$ & 4 & $r_{Q^{PV}}$ & 1\$/MVAR$^2$\\
    $P_{IV}$ & 1000000\$/$pu^2$ & $\eta_c$,$\eta_d$ & 0.92,0.92 & $r_{Q^{CPV}}$ & 1\$/MVAR$^2$\\
    $P_{PC}$ & 10000 \$/MW$^2$ & $SOC$ & [0.2, 0.8] &  $r_{Q^{ES}}$ & 1\$/MVAR$^2$\\
    $B_{cap}$ & 100 MWh & $\beta$ & 90\$/MWh & $J_{c/d}$ & 10e6/5000\$ \\
    $r_{tap}$ & 250 \$/$tap^2$  & $\Delta T$ & 15 min \\
    \midrule
    \multicolumn{4}{c}{\textbf{R} = \text{diag}\{$r_{tap}$,$r_{Q^{PV}}$,$r_{Q^{CPV}}$,$r_{Q^{ES}}$\}} \\
    \end{tabular}%
  \label{tab_para_rcmpc_ctrl}%
   \vspace{-18 pt}
\end{table}%

\subsection{Centralized implementation of proposed RMPC}
\begin{figure}[t]
    \centering
    \includegraphics[width=\linewidth]{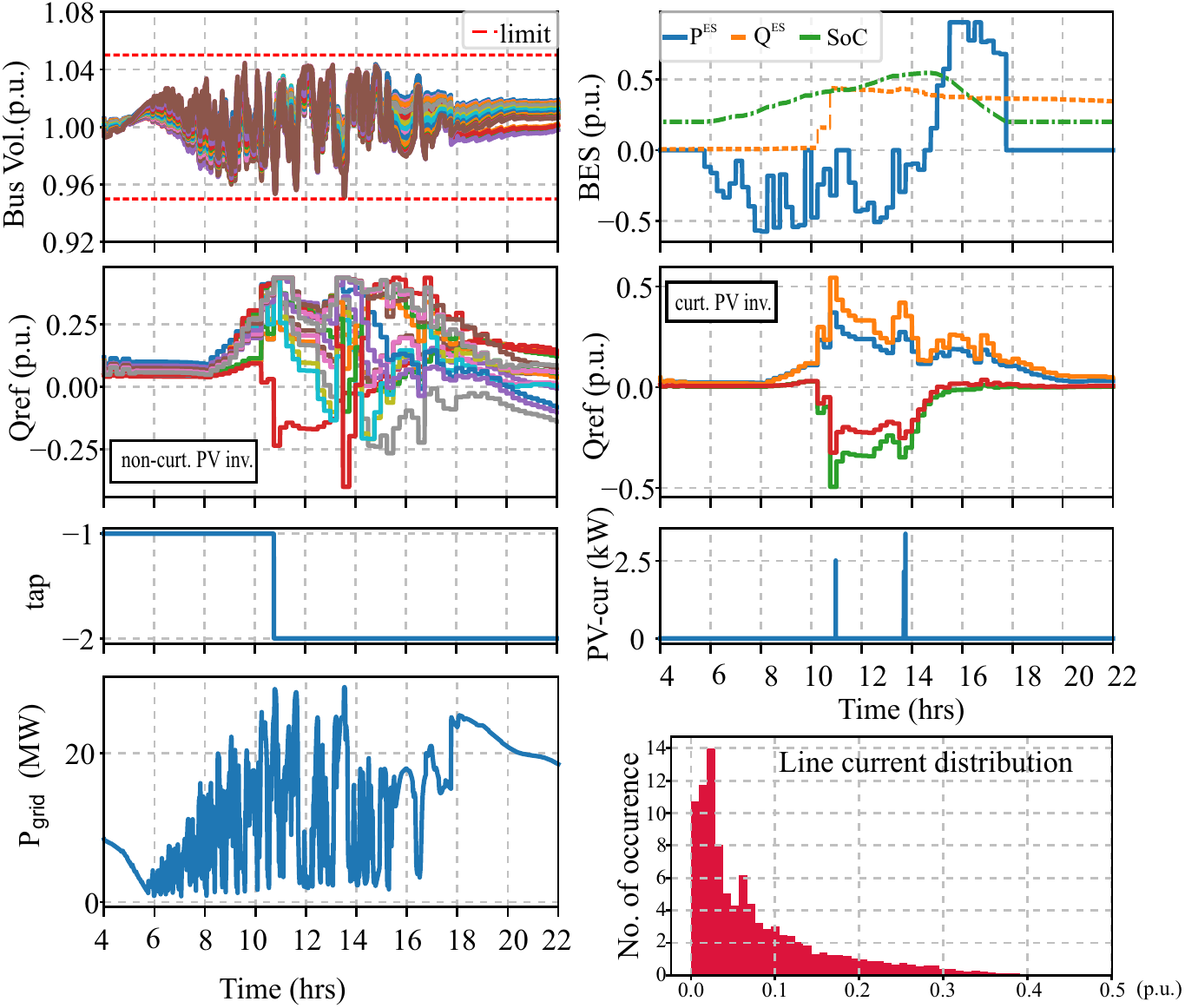}
    \vspace{-20pt}
    \caption{RMPC results in UKGDS network under volatile weather conditions.}
    \label{fig_tecj_eco_V_Var_opt}
    \vspace{-15pt}
\end{figure}

The simulation results with the proposed control scheme are shown in Fig. \ref{fig_tecj_eco_V_Var_opt}. Here, it can be seen that all the node voltages and line currents are completely under the required limit even in the presence of uncertainty in predictions of PVs and loads. The active power from the distribution network started flowing back to the transmission grid at around 6:00 and lasted intermittently till 14:00 in the simulation without any centralized control. However, the proposed RMPC prevented it by allowing BES to absorb the excess generation whenever the aggregate PV production tends to exceed the load demand. The BES also aids to nullify the PV power curtailment. During the period of intermittent reverse power flow, the BES is charged non uniformly which gradually raise the SOC of the battery units. The stored energy in BES is allowed to discharge only when the electricity price exceeds the unit cost of BES discharge. This phenomenon is seen around 15:00 to 18:00 in Fig. \ref{fig_tecj_eco_V_Var_opt}. The reactive power utilization is 169.14 MVAR and the PV curtailment is 0, which is 42.4\% and 100\% lower than the case without BES. The OLTC is utilized only for a single time which was being used 36 times without any centralized control. The reverse power flow is completely eliminated, as shown in the last subplot in Fig. \ref{fig_tecj_eco_V_Var_opt}. 

\subsection{Performance of proposed RMPC at various confidence levels of short-term interval predictions}
All the simulation results presented in the previous subsections were obtained with PV prediction interval computed for 100\% CL. As 100\% CL includes all the samples at the tail of the uncertainty distribution, the control resource utilization required for robust control would be higher, however, it guarantees the ideal performance. The ideal technical requirement for voltage limit violation (VLV), line current violation (LCV), and reverse power flow (RPF) would be absolutely zero. However, minor violations are accepted by DSO in practice. While seeking further saving in control resource utilization, this subsection studies the performance of the proposed RMPC  at lower CL of PV prediction and calculates the resource utilization while compromising violations. The obtained results are tabulated in Table \ref{tab_perf_vs_resource_usage}.
\par 
At 100\% CL of prediction, there is no violation of the voltage and line current limit. Moreover, the reverse power flow is completely ceased with negligible PV curtailment. The reactive and active resource usage accounted for 169.14 MVARh and 69.16 MWh, respectively. At 90\% CL of prediction, the reverse power flow lasted for only 1.33 minutes with only a small peak of 0.51 MW. The total active and reactive resources have been reduced by only 3\% to the previous case; however, the active resource utilization is significantly reduced by 19.12\%. The PV curtailment is still lower, and the tap operation is completely reduced to 0. At lower CL of PV prediction, the total control resources requirement kept reducing; however, the duration and peak value of RPF increased. Depending on the network tolerance for RPF, the suitable value of CL for PV prediction has to be chosen for control in practice.
\begin{table}[!ht]
\vspace{-10pt}
  \centering
  \caption{Performance vs resource utilization of RMPC at various CL of prediction interval}
  \resizebox{\columnwidth}{!}{%
    \begin{tabular}{ccccccccc}
    \toprule
    \textbf{CL of} & \multicolumn{3}{c}{\textbf{duration (min)}} & \textbf{max} & \textbf{Q} & \textbf{P} & \textbf{curtail} & \textbf{tap} \\
    \textbf{PI} & \textbf{VLV} & \textbf{LCV} & \textbf{RPF} & \textbf{RPF(MW)} & \textbf{MVARh} & \textbf{MWh} & \textbf{MWh} &  \\
    \midrule
    100\% & 0     & 0     & 0     & 0     & 169.14 & 69.16 & 0.0001 & 1 \\
    \midrule
    90\%  & 0     & 0     & 1.33  & 0.51  & 175.09 & 55.93 & 0.0023 & 0 \\
    \midrule
    80\%  & 0     & 0     & 6.23  & 2.16  & 145.52 & 48.06 & 0.0073 & 0 \\
    \midrule
    70\%  & 0     & 0     & 14.2  & 3.85  & 147.3 & 40.63 & 0.0267 & 0 \\
    \bottomrule
    \end{tabular}}
  \label{tab_perf_vs_resource_usage}%
  \vspace{-11pt}
\end{table}%

\section{Conclusions}
This paper formulates an RMPC for achieving real-time techno-economic operation of the distribution network considering the uncertainties of PVs and loads. Distributed resources such as battery energy storage (BES) and curtailable/non-curtailable PVs along with tap-changers are utilized as control resources in the proposed controller. The proposed RMPC was tested for closed-loop control of the UKGDS network. Even at 80\% PV penetrating, the RMPC was able to achieve regulation of voltage, line current, reverse power flow, and economic operation in a robust manner considering the prediction interval forecast of PVs and loads. Specifically, the RMPC was able to utilize BES for reducing the curtailment of curtailable PVs and discharging/charging based on the time-varying electricity price of the grid. 
Furthermore, the performance of RMPC was found to be dependent on the prediction interval (PI) of PVs. Case studies conducted at various confidence levels (CL) of the PI revealed that the control resource utilization gets reduced at lower CL, however, violation mainly in the reverse power flow increases. The case studies infer to select a suitable value of CL for PI forecast of PVs which trade-off acceptable violation limits and control resources.

\bibliographystyle{IEEEtran}
\bibliography{references1.bib}
\end{document}